\documentclass[aps,prd,twocolumn,showpacs,superscriptaddress,nofootinbib,preprintnumbers]{revtex4-2}
\usepackage[utf8]{inputenc}
\usepackage{amssymb}
\usepackage{hhline}
\usepackage{amsmath}
\usepackage{mathtools}
\usepackage[dvipsnames]{xcolor}
\usepackage{multirow,tabularx}
\usepackage{graphicx}
\usepackage{natbib}

\usepackage{ulem}

\usepackage{xspace}
\usepackage{xstring}
\usepackage{parskip}
\allowdisplaybreaks
\usepackage{braket}

\newcommand{\gsim}{\raisebox{-0.13cm}{~\shortstack{$>$ \\[-0.07cm]
      $\sim$}}~}
\newcommand{\lsim}{\raisebox{-0.13cm}{~\shortstack{$<$ \\[-0.07cm]
      $\sim$}}~}

\newrobustcmd{\pea}[1]{%
	\emph{#1}\textbf{\ \ \ ---}
}

\usepackage[colorlinks=true,citecolor=blue,urlcolor=blue]{hyperref}
\hypersetup{colorlinks=true,citecolor=romared,linkcolor=romared,urlcolor=romared}
\definecolor{romared}{RGB}{142,0,28}

\definecolor{tabblue}{RGB}{31, 119, 180}
\definecolor{darkblue}{RGB}{0, 0, 120}
\definecolor{tabred}{RGB}{214, 39, 40}
\definecolor{tabgreen}{RGB}{44, 160, 44}
\definecolor{tabgray}{RGB}{100, 100, 100}
\begin{document}
\preprint{\hbox{UTWI-09-2025, NORDITA-2025-015}}
\title{Can a Breakdown of Hawking Evaporation Open a New\\Mass Window for Primordial Black Holes as Dark Matter?}


\author{Gabriele Montefalcone}
\affiliation{Texas Center for Cosmology and Astroparticle Physics, Weinberg Institute for Theoretical
Physics, Department of Physics, University of Texas at Austin, Austin, TX 78712, USA}

\author{Dan Hooper}
\affiliation{Department of Physics, University of Wisconsin, Madison, WI, USA}
\affiliation{Wisconsin IceCube Particle Astrophysics Center (WIPAC),
University of Wisconsin, Madison, WI, USA}

\author{Katherine Freese}
\affiliation{Texas Center for Cosmology and Astroparticle Physics, Weinberg Institute for Theoretical
Physics, Department of Physics, University of Texas at Austin, Austin, TX 78712, USA}
\affiliation{Oskar Klein Centre, Department of Physics, Stockholm University, 10691 Stockholm, SE}
\affiliation{Nordita, KTH Royal Institute of Technology and Stockholm University, 10691 Stockholm, SE}

\author{Chris Kelso}
\affiliation{Department of Physics, University of North Florida, 1 UNF Dr, Jacksonville, FL 32224, USA}

\author{Florian K{\"u}hnel}
\affiliation{Arnold Sommerfeld Center, Ludwig-Maximilians-Universit{\"a}t, Theresienstr.~37, 80333 M{\"u}nchen, Germany}
\affiliation{Max-Planck-Institut f{\"u}r Physik, Boltzmannstr.~8, 85748 Garching, Germany}

\author{Pearl Sandick}
\affiliation{Department of Physics and Astronomy, University of Utah, Salt Lake City, UT, 84112, USA}
\affiliation{Laboratoire Univers et Particules de Montpellier, CNRS \& Universit{\'e} de Montpellier, France}

\begin{abstract}

Semi-classical Hawking evaporation is expected to break down at some point in a black hole’s evolution as the effects of quantum gravity become important.  In particular, it has been argued that the so-called memory-burden effect could cause black holes to become stabilized by the information that they carry, thereby suppressing the rate at which they undergo Hawking evaporation. It has furthermore been suggested that this opens a new mass window, between $10^{4}\,{\rm g} \lsim M \lsim 10^{10}\,{\rm g}$, over which primordial black holes could constitute the dark matter of our Universe. We show for the first time that this is true only if the transition from the semi-classical phase of a black hole to its memory-burdened phase is practically instantaneous. If this transition is instead more continuous, Hawking evaporation will persist at relevant levels throughout the eras of Big Bang Nucleosynthesis and recombination, leading to stringent constraints which rule out the possibility that black holes lighter than $\sim 4 \times 10^{16}\,{\rm g}$ could make up all or most of the dark matter. More broadly, our analysis demonstrates that even if departures from the semi-classical Hawking evaporation occur as proposed, they must be both drastic and abrupt to open viable new mass windows for primordial black hole dark matter.\\[-1mm]
\end{abstract}

\maketitle


\section{Introduction}
 \noindent It has long been speculated that a population of primordial black holes (PBHs) may have formed in the early Universe~\cite{Zeldovich:1967lct, Hawking:1971ei} and could constitute some or all of our Universe's dark matter~\cite{Chapline:1975ojl, Meszaros:1975ef}. Hawking famously argued that black holes should radiate particles from their horizons, causing them to lose mass and evaporate. In the standard semi-classical picture, black holes without appreciable electric charge or angular momentum lose mass through this process at the following rate~\cite{Hawking:1975vcx}:
\begin{align}
\label{eq:SC}
    \frac{{\rm d}M}{{\rm d}t}\bigg|_{\rm SC}
        &=
            -\mspace{2mu}\frac{\mathcal{G}\,g_{\star, H}(T_{\rm BH}) M_{\rm Pl}^{4}}{30720 \pi M^{2}}\\
        &\approx
            -\,8.2 \times 10^{6}\,{\rm g/s} \times\! \bigg(\frac{g_{\star, H}(T_{\rm BH})}{108}\bigg)\!\bigg(\frac{10^{10}\,{\rm g}}{M}\bigg)^{\!2}, \nonumber
\end{align}
where $\mathcal{G} \approx 3.8$ is the appropriate greybody factor, $T_{\rm BH} = M^{2}_{\rm Pl}/(8\pi M)$ is the temperature of the black hole, $M_{\rm Pl} \approx 2.2 \times 10^{-5}\,{\rm g}$ is the Planck mass, and $g_{\star, H}$ counts the weighted degrees-of-freedom of the particle species lighter than $T_{\rm BH}$, normalized to 1 per fermionic degree of freedom~\cite{MacGibbon:1990zk, MacGibbon:1991tj}. Through this process of semi-classical Hawking evaporation, a black hole with an initial mass, $M_{i}$, will evaporate completely in a time given by
\begin{align}
    t_{\rm evap}
        &=
            \frac{30720 \pi}{\mathcal{G} M^{4}_{\rm Pl}} \int_{0}^{M_{i}} \frac{M^{2}\, {\rm d}M}{g_{\star, H}(T_{\rm BH})}\\[2mm]
        &\approx
            400\,{\rm s} \times \bigg(\frac{M_{i}}{10^{10}\,{\rm g}}\bigg)^{\!3} \bigg(\frac{108}{\big\langle g_{\star, H}(T_{\rm BH})\big\rangle}\bigg)\mspace{1.5mu} , \nonumber 
\end{align}
where $\langle g_{\star, H} \rangle$ is the appropriate average of $g_{\star, H}$ over the lifetime of the black hole. From this, we conclude that black holes lighter than $\sim 5 \times 10^{14}\,{\rm g}$ will evaporate in less than the age the Universe. Black holes with masses in the range of $\sim 5 \times 10^{14}$ to $\sim 10^{17}\,{\rm g}$ would produce detectable fluxes of gamma rays and other forms of Hawking radiation if they were to make up all of our Universe's dark matter~\cite{Coogan:2020tuf, Boudaud:2018hqb, PhysRevD.101.123514, Keith:2021guq, Dasgupta:2019cae}. On the other hand, constraints from gravitational microlensing surveys appear to have ruled out the possibility that black holes (or other compact objects) heavier than $\sim 5 \times 10^{21}\,{\rm g}$ could constitute all of the dark matter~\cite{Macho:2000nvd, EROS-2:2006ryy, Griest:2013aaa, Oguri:2017ock, Niikura:2019kqi, Croon:2020ouk} (for other perspectives on this, see Refs.~\cite{Carr:2023tpt,Garcia-Bellido:2024yaz}). In light of these constraints, the only mass range in which PBHs could constitute all or most of our Universe's dark matter is that of $10^{17}\,{\rm g} \lsim M_{\rm BH} \lsim 5 \times 10^{21}\,{\rm g}$.

Recent theoretical developments have motivated us to revisit some of these conclusions. In particular, it has been suggested that the so-called memory-burden effect~\cite{Dvali:2018xpy, Dvali:2018ytn} could cause systems to be stabilized by the information they carry~\cite{Dvali:2020wft}. This effect, originally established in large-N quantum systems, has been proposed to apply universally to high-entropy systems, such as black holes. While this analogy remains speculative, it is physically well-motivated and has generated significant interest in the community. In fact, when applied to black holes, the memory-burden effect could significantly suppress the process of Hawking evaporation, thereby opening a mass range over which very light PBHs would be cosmologically long-lived, and could constitute the dark matter of our Universe~\cite{Dvali:2020wft, Alexandre:2024nuo, Dvali:2024hsb, Thoss:2024hsr},  as well as yielding distinct observational signatures~\cite{Zantedeschi:2024ram,Kohri:2024qpd,Barman:2024iht,Jiang:2024aju,Bhaumik:2024qzd}.

While we take no position on the validity of the memory-burden mechanism itself, analyzing its proposed implications for PBH dark matter is a useful testbed to assess more generally how a breakdown of Hawking evaporation impacts cosmological observables. In this work, we revisit  the constraints on light PBHs, taking into account the effects of memory burden. In previous works, it was claimed that this mechanism would make it possible for PBHs with masses in the range of $\sim 10^{4}$ -- $10^{10}\,{\rm g}$ to constitute all of the dark matter. In contrast, we find here that measurements of the cosmic microwave background (CMB) and the primordial light element abundances severely constrain the abundances of PBHs lighter than $\sim 4 \times 10^{16}\,{\rm g}$. Unless memory burden sets in almost instantaneously in a black hole's evolution, effectively preventing any Hawking radiation from being produced, we find that PBHs in this mass range cannot make up a significant fraction of the dark matter. More generally, this demonstrates that any mechanism suppressing Hawking evaporation faces severe cosmological constraints unless the suppression is both extreme and abrupt.
\vspace{1mm}

\section{Memory Burden and Primordial Black Holes}
\noindent In the existing literature, the impact of the memory-burden effect on black hole evaporation has generally been parameterized in terms a step-like function~\cite{Dvali:2020wft, Alexandre:2024nuo, Dvali:2024hsb}:
\begin{align}
    &\qquad
        \frac{{\rm d}M}{{\rm d}t} =
        \begin{dcases}
            \frac{{\rm d} M}{{\rm d} t}\Big|_{\rm SC}
                & \text{if } M \geq q M_{i}\\[1mm]
            \frac{{\rm d} M}{{\rm d} t}\Big|_{\rm MB}
                & \text{if } M < q M_{i}\, ,
        \end{dcases} \quad & 
\label{eq:stepMB}
\end{align}
where $M_{i}$ is the initial mass of the black hole and $q$ is the mass fraction at which the transition to the memory-burdened phase takes place. ${\rm d}M/{\rm d}t\,|_{\rm SC}$ is the standard semi-classical evaporation rate, as given in Eq.~\eqref{eq:SC}. The evaporation rate in the memory-burden phase is constant and given by
\begin{align}
    \frac{{\rm d}M}{{\rm d}t}\bigg|_{\rm MB}
        =
            \frac{1}{\Tilde{S}(q M_{i})^k} \cdot 
            \frac{{\rm d}M}{{\rm d}t} \big(qM_{i}\big)\bigg|_{\rm SC},
\label{eq:MB}
\end{align}
where $k$ is expected to be an integer (with $1\leq k \leq 3$)~\cite{Dvali:2024hsb}, and $\Tilde{S}$ is the dimensionless entropy of the black hole, defined as~\cite{Bekenstein:1973ur}
\begin{equation}
    \Tilde{S}
        \equiv
            \frac{S}{k_{\rm B}}
        = 
            \frac{\pi r_{\rm g}^{2}}{\hbar G\mspace{1.5mu}k_{\rm B}}   
        \approx
            2.6\times 10^{30} \times \!\left(\frac{M}{10^{10}\,{\rm g}}\right)^{\!2}
            ,
\end{equation} 
where $k_{\rm B}$ is the Boltzmann constant and $r_{\rm g}$ is the Schwarzschild radius of the black hole. Throughout this work, we will adopt $k=2$, as suggested by numerical studies and theoretical considerations~\cite{Dvali:2020wft,Dvali:2024hsb,Dvali:2017nis}.
For results adopting other values of $k$, we direct the reader to Appendix~\ref{sec:s2}, specifically Fig.~\ref{fig:results_4}.

According to this step-like parameterization, black holes evaporate in the normal semi-classical fashion until they have lost a fraction, $1-q$, of their mass, after which any Hawking radiation is suppressed by the very large factor of $\Tilde{S}^{k}$~\cite{Dvali:2018xpy, Dvali:2018ytn, Dvali:2020wft}. This makes it possible that black holes much lighter than $\sim 5 \times 10^{14}\,{\rm g}$ could be stabilized by the effect of memory burden, allowing them to serve as a candidate for dark matter. To evade constraints from the measured primordial element abundances, these black holes would have to be lighter than $\sim 10^{10}\,{\rm g}$~\cite{Keith:2020jww,Haque:2024eyh}, allowing them to enter into the memory-burdened phase before the onset of Big Bang Nucleosynthesis (BBN). Whereas a $\sim 10^{10}\,{\rm g}$ ($\sim 10^{6}\,{\rm g}$) black hole would evaporate semi-classically in only $\sim 400\,{\rm s}$ ($\sim 4 \times 10^{-10}\,{\rm s}$), the effect of memory burden would extend the lifetime of these objects to $t_{\rm evap} \sim 10^{60}\,{\rm s}$ ($t_{\rm evap} \sim 10^{32}\,{\rm s}$), for $k=2$ and $q=0.5$, for example. Adopting the step-like parameterization described in Eq.~\eqref{eq:stepMB}, any black holes in the mass range of $\sim 10^{4}$ -- $10^{10}\,{\rm g}$ would be cosmologically long-lived and yet enter into the memory-burdened phase prior to the era of BBN. Such black holes would be consistent with the measured light element abundances, as well as with constraints from gamma-ray and cosmic-ray measurements.

\begin{figure}
    \centering
    \includegraphics[width=0.95\linewidth, height=.833333\linewidth]{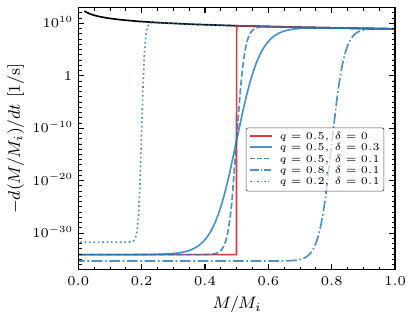}
    \caption{
        The evaporation rate of a black hole with an initial mass of $M_{i} = 10^{6}\,\mathrm{g}$ (corresponding to a semi-classical evaporation time of $t_{\rm evap} \sim 4 \times 10^{-10}\,\mathrm{s}$) as a function of the remaining mass fraction, $M/M_{i}$. The black curve represents the semi-classical approximation (see Eq.~\eqref{eq:SC}), while the red curve depicts the step-like memory burden approximation,  (see Eq.~\eqref{eq:stepMB}), corresponding to the case of $q = 0.5$ and $\delta = 0$. The blue curves show several examples of a smooth transition between the semi-classical and memory-burdened phases, for selected choices of $q$ and $\delta$ (see Eq.~\eqref{eq:smooth_MB}). Here, we have adopted $k=2$.
    }
    \label{fig:dmdt}
\end{figure}

A central point of this work is that this conclusion is highly sensitive to the instantaneous nature of the transition between the semi-classical and memory-burdened phases, as described in Eq.~\eqref{eq:stepMB}. To allow for a more continuous and realistic transition between these two phases, we introduce the following parameterization:
\begin{equation}
    \frac{{\rm d}M}{{\rm d}t}
        =
            \left(\frac{{\rm d}M}{{\rm d}t}\bigg|_{\rm SC}\right)^{\!h}\times \left(\frac{{\rm d}M}{{\rm d}t}\bigg|_{\rm MB}\right)^{\!(1-h)}, \label{eq:smooth_MB}
\end{equation}
where
\begin{align}
    h(M)
        &\equiv
            \frac{1}{2}\bigg(1+\tanh\left[\frac{M-q M_{i}}{\delta \left(q\,M_{i}/2\right)}\right]\bigg)\, .
\end{align}
The new parameter, $\delta$, sets how quickly the transition takes place between the two phases, and in the limit of $\delta \rightarrow 0$, we recover the step-like behavior described in Eq.~\eqref{eq:stepMB}. For $M \gg q \,M_{i} \,(1+\delta/2)$, the evaporation rate reduces to semi-classical prediction, while at $M \ll q \,M_{i} \,(1+\delta/2)$, the evaporation rate reduces to the behavior predicted in the memory-burdened phase. In Fig.~\ref{fig:dmdt}, we compare the step-like evaporation rate to that of our continuous parameterization, for several choices of $q$ and $\delta$, and for the case of a black hole with an initial mass of $M_{i} = 10^{6}\,\mathrm{g}$ (corresponding to a semi-classical evaporation time of $t_{\rm evap} \sim 4\times 10^{-10}\,\mathrm{s}$). Note that the evaporation rate in the $M \ll M_i$ regime is sensitive to the value of $q$, as follows from Eq.~\eqref{eq:MB}.
\vspace{1mm}
%

\section{Constraints from BBN and the CMB}
\begin{figure}
    \centering
    \includegraphics[width=0.95\linewidth, height=.833333\linewidth]{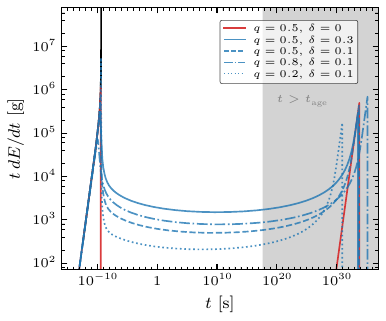}
    \caption{
        The time profile of the Hawking radiation from a black hole with an initial mass of $M_{i} =10^{6}\,\mathrm{g}$ (corresponding to a semi-classical evaporation time of $t_{\rm evap} \sim 4\times 10^{-10}\,\mathrm{s}$). The black curve represents the semi-classical approximation (see Eq.~\eqref{eq:SC}), while the red curve depicts the step-like memory burden approximation (see Eq.~\eqref{eq:stepMB}), corresponding to $q = 0.5$ and $\delta = 0$. The blue curves represent the profile for the continuous parameterization of Eq.~\eqref{eq:smooth_MB}, for several values of $q$ and $\delta$. Here, we have adopted $k=2$.
    }
\label{fig:time_profs}
\end{figure}

\noindent Black holes which transition continuously between their semi-classical and memory-burdened phases remain in this transitory phase for quite some time, during which they produce significant fluxes of Hawking radiation. This is illustrated in Fig.~\ref{fig:time_profs}, where we plot the time profile of the Hawking radiation from a black hole with an initial mass of $M_{i} = 10^{6}\,\mathrm{g}$. In the semi-classical approximation, the black hole disappears entirely after $\sim 4 \times 10^{-10}\,{\rm s}$, and thus has no impact on the primordial light element abundances or on the CMB. In the step-like memory burden approximation, Hawking evaporation is constant and entirely negligible between $10^{-10}\,{\rm s}$ and $10^{30}\,{\rm s}$, again consistent with all data. If there is a continuous transition between these phases, however, a significant flux of Hawking radiation will be produced throughout this window, allowing us to place stringent constraints on PBHs in this mass range.

The injection of energetic particles in the era between recombination and reionization can heat and ionize neutral hydrogen, thereby altering the thermal history of the Universe. This, in turn, can impact the measured characteristics of the CMB, including its spectral shape and distribution of temperature anisotropies~\cite{Padmanabhan:2005es, Slatyer:2009yq, Chluba:2011hw}. Similarly, the injection of energetic particles during BBN can alter the primordial proton-to-neutron ratio in way that increases the resulting helium abundance. Such energy injection can also break apart helium nuclei and create deuterium, through the processes of photodissociation and hadrodissociation~\cite{Jedamzik:2009uy, Carr:2009jm, Pospelov:2010hj}.

\begin{figure}
    \centering
    \includegraphics[width=0.95\linewidth, height=0.83333\linewidth]{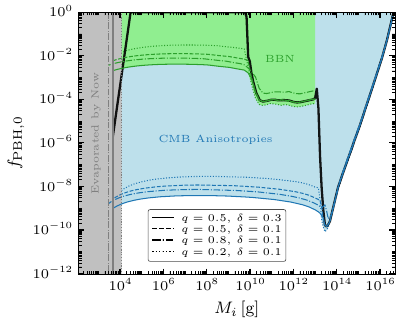}
    \caption{
        Upper limits on the abundance of primordial black holes as a function of their initial mass, shown for several values of $q$ and $\delta$. Constraints are derived from both the primordial light element abundances (green) and CMB anisotropies (blue). The gray region corresponds to primordial black holes that would have fully evaporated by now. The quantity, $f_{{\rm PBH},0}$, is the initial fraction of the dark matter that consists of primordial black holes. The thick solid black line highlights the bounds for a step-like memory-burden transition $(q = 0.5,\,\delta = 0)$, which would leave a relatively large mass window unconstrained, $ 10^{4} \, {\rm g} \lsim M_i \lsim 10^{10} \, {\rm g}$~\cite{Alexandre:2024nuo, Dvali:2024hsb, Thoss:2024hsr}. As we have shown here, however, if there is a continuous transition between the semi-classical and memory-burdened phases of a black hole's evolution ($\delta>0$), this window closes due to the strong constraints from both BBN and the CMB. Here, we have adopted $k=2$.
    } 
    \label{fig:results}
\end{figure}

We calculate our constraints from the primordial light element abundances and from the CMB following the procedures described in Refs.~\cite{Keith:2020jww} (based on results presented in Ref.~\cite{Kawasaki:2017bqm}) and~\cite{Acharya:2019uba} (see also, Refs.~\cite{Slatyer:2016qyl, Poulin:2016anj} and Appendix~\ref{sec:s1} for more details), respectively. Our main results are shown in Fig.~\ref{fig:results}. In this figure, we show the regions of parameter space that are ruled out by these measurements, for various choices of $q$ and $\delta$. Across all of the parameter space shown, PBHs can make up only a very small fraction of the total dark matter. For further details, demonstrating the robustness of our conclusions across different values of $q$, $\delta$, and $k$, see Appendix~\ref{sec:s2}.
\vspace{1mm}

\section{Early Onset Memory Burden}
\noindent Throughout this work, we have made the common assumption that the memory-burdened phase begins after a black hole has lost an order one fraction of its initial mass, $q \sim 0.1$ -- $0.9$~\cite{Dvali:2020wft, Alexandre:2024nuo, Dvali:2024hsb, Thoss:2024hsr}. If this is not in fact the case, then the constraints we have derived here will be modified. In particular, we have shown in Appendix~\ref{sec:s2} that a new mass window for PBH dark matter would indeed appear if the memory-burdened phase begins extremely early in the evaporation of a black hole (see Fig.~\ref{fig:results_2}). Quantitatively, this would require $(1-q) \lsim 10^{-10}$.

It would perhaps be surprising to learn that black holes enter into the memory-burdened phase after losing only a tiny fraction of their initial mass. Such a picture could be motivated, however, within the context of the original model of Ref.~\cite{Dvali:2020wft} (see also \cite{Dvali:2017ktv, Dvali:2018tqi, Dvali:2018xpy, Dvali:2018ytn,Michel:2023ydf,Thoss:2024hsr,Dvali:2024hsb}), in which there is an attractive interaction between the memory burden modes. For a critical occupation, $\braket{\hat{n}_{0}} = M_{0} / r_{\rm g}^{-1}$, the positive free energy gaps are precisely canceled by the negative interaction energy, causing the modes to become effectively gapless~\cite{Dvali:2017ktv, Dvali:2018tqi, Dvali:2018xpy, Dvali:2018ytn, Dvali:2020wft}. From this perspective, even a small change in the mass of a black hole could effectively stabilize it and prevent it from appreciably evaporating~\cite{Dvali:2018xpy, Dvali:2018ytn, Dvali:2020wft, Dvali:2021tez, Dvali:2023xfz, Dvali:2024hsb}. In such a case, the fraction of a black hole's mass that is lost before it enters the memory-burdened phase scales as an inverse power of the black hole entropy, generically yielding values of $(1-q) \ll 10^{-10}$. As a result, the constraints from BBN and the CMB can be straightforwardly evaded in such a scenario. 
\vspace{1mm}

\section{Summary and Conclusions}

\noindent In this work, we have revisited the constraints on the abundance of primordial black holes from the cosmic microwave background and the primordial light element abundances, considering the possible impact of the memory-burden effect. The central idea of memory burden is that an object can be stabilized by the information it carries. This can, for example, suppress the rate at which black holes produce Hawking radiation. It has been previously suggested that this could open a new range of masses, $M \sim 10^{4}$ -- $10^{10}\,{\rm g}$, over which primordial black holes could make up the dark matter~\cite{Dvali:2020wft, Alexandre:2024nuo, Dvali:2024hsb, Thoss:2024hsr}. We have shown here for the first time, however, that this is the case only if the transition between the semi-classical and memory-burdened phases of black hole evaporation is nearly instantaneous. If this instead occurs more continuously, then Hawking evaporation will take place at relevant levels throughout the eras of Big Bang Nucleosynthesis and recombination, allowing us to rule out the possibility that black holes lighter than $\sim 4 \times 10^{16}\,{\rm g}$ could make up all or most of the dark matter.

These results rely on the assumption that the memory-burdened phase begins after a black hole has lost an order-one fraction of its initial mass, $q \sim (0.1$ -- $0.9)$. If the onset of the memory-burden effect instead takes place almost immediately after the formation of a black hole, then our constraints could be weakened. Memory-burden stabilized black holes could constitute all of the dark matter if $(1-q) \lsim 10^{-10}$. 

In summary, independent of whether the memory-burden effect is ultimately realized in nature, we have shown here that this mechanism cannot open the claimed mass window for PBH dark matter without an almost instantaneous transition to the suppressed evaporation. More broadly, our analysis establishes that any modification to the semi-classical Hawking evaporation -- whether due to memory burden or other physical mechanisms -- must be both drastic and abrupt to evade the stringent cosmological constraints from CMB and BBN.
\bigskip

\section*{Acknowledgements}
\noindent We thank Michael Zantedeschi and Sebastian Zell for early discussions on the parameterization of the memory burden effect. For related work see Ref.~\cite{Dvali:2025ktz}. D.\,H.~is supported by the Office of the Vice Chancellor for Research at the University of Wisconsin--Madison, with funding from the Wisconsin Alumni Research Foundation. K.\,F.~and G.\,M.~are grateful for support from the Jeff \& Gail Kodosky Endowed Chair in Physics held by~K.\,F.~at the University of Texas at Austin. K.\,F.~and G.\,M.~also acknowledge funding from the US~Department of Energy under Grant~\mbox{DE-SC-0022021}, as well as the Swedish Research Council (Contract No.~638-2013-8993). G.\,M.~is also supported by the Continuing Fellowship of the Graduate School of the College of Natural Sciences at the University of Texas at Austin. The work of P.\,S.~is supported in part by the National Science Foundation grant PHY-2412834. The work of C.\,K.~is supported in part by the U.S.~Department of Energy, Office of Science, Office of High Energy Physics under Award Number DE-SC0024693.

\appendix
\section{Methodology}\label{sec:s1}
In Ref.~\cite{Kawasaki:2017bqm}, the authors employed a sophisticated code to study the effects of long-lived particles on the primordial light element abundances. These results were adapted in Ref.~\cite{Keith:2020jww} to derive constraints on the presence of PBHs, assuming their semi-classical evaporation.\footnotetext{For more recent BBN constraints on semi-classically evaporating PBHs, see Ref.~\cite{Carr:2020gox}; here we conservatively follow the approach of Ref.~\cite{Keith:2020jww}.}  In this work, we have followed this same procedure, modified to account for the effects of memory burden on the time and temperature profile of the Hawking radiation. 

For each set of parameters, we first derived a constraint based purely on the semi-classical phase of the black holes' evaporation (defined as the time window over which $M > q M_{i}$). Following Ref.~\cite{Keith:2020jww}, we found the time at which the median unit of energy (in the photodisassociation era of BBN, $T \lsim 0.4\,{\rm keV}$~\cite{Kawasaki:1994af}) or the median meson (in the hadrodisassociation era, $T \gsim 0.4\,{\rm keV}$) is produced through black hole evaporation, and then mapped this result onto the corresponding lifetime of a decaying particle~\cite{Kawasaki:1994af}. We further calculated the average energy of the particles that were radiated from the black holes during this period and mapped this result onto a mass for the corresponding decaying particle species. 

For black holes in the memory-burdened phase, we have mapped the time ane temperature profile onto that of a decaying particle species with a lifetime, $\tau$, in the range of $10 \times t_{\rm MB}$ and $10^{-3} \times t_{\rm evap}$, where $t_{\rm MB}$ is the time at which $M=q M_{i}$. For each lifetime, we calculated the total energy radiated from black holes between $t= 10^{-3}\,\tau$ and $10^{2}\,\tau$. If this weighted quantity of energy injection exceeded the constraints presented in Refs.~\cite{Keith:2020jww, Kawasaki:2017bqm} for any of the lifetimes in the stated range, we consider the scenario in question to be ruled out. Note that we have taken the temperature of the black hole to be constant during the memory-burdened phase, with $T_{\rm BH} = M^{2}_{\rm Pl}/(8 \pi q M_{i})$.

To derive constraints based on the CMB, we have made use the results presented in Ref.~\cite{Acharya:2019uba}, which considered the case of decaying particles (see also, Refs.~\cite{Slatyer:2016qyl, Poulin:2016anj}). We applied these results in close analogy to how we applied the results of Refs.~\cite{Keith:2020jww,Kawasaki:2017bqm}, again modifying the approach to account for the effects of memory burden on the time and temperature profiles of black hole evaporation. Note that the constraints from the CMB are largely insensitive to the mass of the decaying particle species~\cite{Slatyer:2016qyl}, so we simplified our procedure by taking the weakest constraint from Fig.~9 of Ref.~\cite{Acharya:2019uba}, regardless of the black hole's temperature. This approximation is conservative and in some cases could potentially weaken our constraints by as much as an order of magnitude.

\section{Results for Other Choices of $q$, $\delta$, and $k$}\label{sec:s2}

\begin{figure*}
\vspace{-.5cm}
    \centering
\includegraphics[width=1\linewidth]{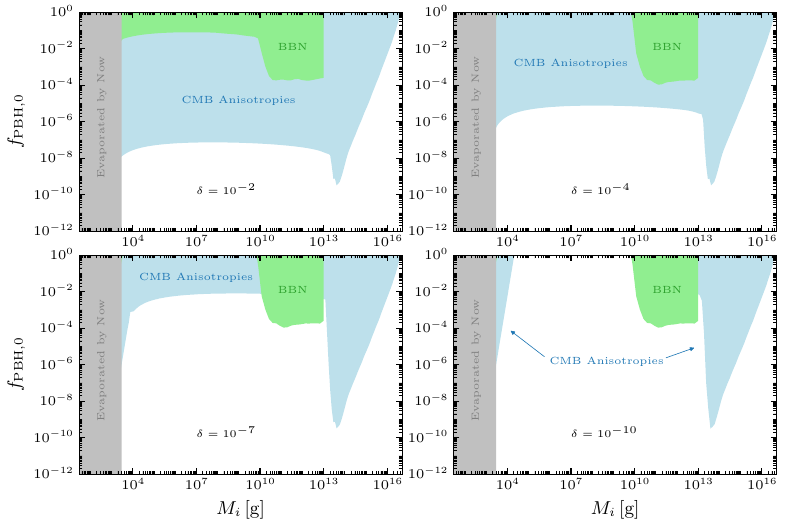}
    \caption{Upper limits on the abundance of primordial black holes as a function of their initial mass, shown for four small values of $\delta$. In the limit $\delta\rightarrow 0$, we recover the results of the step-like memory burden approximation. Constraints are derived from both the primordial light element abundances (green) and CMB anisotropies (blue), highlighting how a continuous transition to the memory-burden phase affects these bounds. The gray region corresponds to primordial black holes that would have fully evaporated within the age of the Universe. Here, we have adopted $q=0.8$ and $k=2$, with $f_{{\rm PBH},0}$ denoting the initial fraction of the dark matter that consists
of primordial black holes.}
    \label{fig:results_3}
\end{figure*}

In the main body of this manuscript, we presented results for a small selection of $q$ and $\delta$. Here, we show results for other choices of these parameters. In Fig.~\ref{fig:results_3}, our constraints are given for the case of $q=0.8$ and $k=2$, and for four small values of $\delta$. In the limit of $\delta \rightarrow 0$, we recover the results of the step-like approximation. 

\begin{figure*}
    \centering
    \includegraphics[width=.5\linewidth, height=0.416666\linewidth]{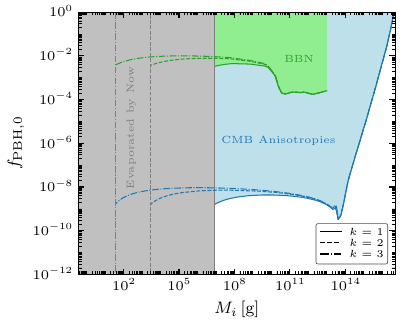}
    \caption{As in Fig.~\ref{fig:results_3}, but for three values of $k$. From this comparison, we see that this parameter choice has little impact on the resulting bounds.  Here, we have adopted $q=0.8$ and $\delta=0.1$.}
    \label{fig:results_4}
\end{figure*}

Throughout this work, we have adopted $k=2$, as suggested by numerical studies and theoretical considerations~\cite{Dvali:2020wft,Dvali:2024hsb,Dvali:2017nis}. In Fig.~\ref{fig:results_4}, we compare these results to those obtained for $k=1$ and $k=3$. From this, we see that this parameter has little impact on our constraints.\vspace{0.5cm}

If the parameter, $q$, is sufficiently close to unity, black holes will end their semi-classical phase after losing only a tiny fraction of their initial mass, becoming quickly stabilized by the memory-burden effect. If this is the case, then the constraints from BBN and the CMB will be much weaker than what we have found in the case of $q\sim 0.1-0.9$.

\begin{figure*}
    \centering
    \includegraphics[width=1\linewidth]{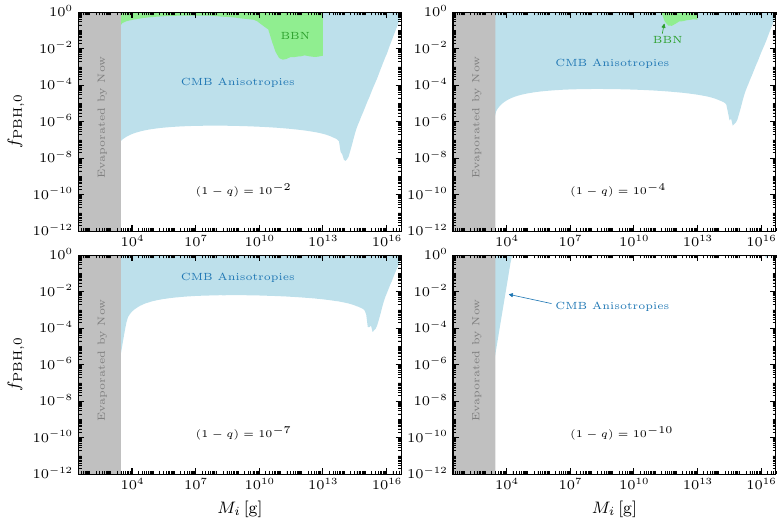}
    \caption{As in Fig.~\ref{fig:results_3}, but for four small values of $(1-q)$. Here, we have adopted $k=2$ and $\delta/(1-q)=0.1$.  
    A new mass window over which primordial black holes can constitute the dark matter of our Universe appears only if  $(1-q) \lsim 10^{-10}$.}
    \label{fig:results_2}
\end{figure*}

We illustrate this in Fig.~\ref{fig:results_2}, where we plot the constraints on PBHs for four small values of $(1-q)$. Here, we have adopted $k=2$ and $\delta/(1-q) =0.1$. Only for $(1-q) \lsim 10^{-10}$ does the effect of memory burden open a new mass window for PBHs to make up the dark matter of our Universe.  

\bibliographystyle{apsrev4-1}
\bibliography{Refs}

@article{Haque:2024eyh,
    author = "Haque, Md Riajul and Maity, Suvashis and Maity, Debaprasad and Mambrini, Yann",
    title = "{Quantum effects on the evaporation of PBHs: contributions to dark matter}",
    eprint = "2404.16815",
    archivePrefix = "arXiv",
    primaryClass = "hep-ph",
    doi = "10.1088/1475-7516/2024/07/002",
    journal = "JCAP",
    volume = "07",
    pages = "002",
    year = "2024"
}

@article{Dvali:2017nis,
    author = "Dvali, Gia",
    title = "{Area law microstate entropy from criticality and spherical symmetry}",
    eprint = "1712.02233",
    archivePrefix = "arXiv",
    primaryClass = "hep-th",
    doi = "10.1103/PhysRevD.97.105005",
    journal = "Phys. Rev. D",
    volume = "97",
    number = "10",
    pages = "105005",
    year = "2018"
}

@article{Carr:2023tpt,
    author = "Carr, Bernard and Clesse, Sebastien and Garcia-Bellido, Juan and Hawkins, Michael and Kuhnel, Florian",
    title = "{Observational evidence for primordial black holes: A positivist perspective}",
    eprint = "2306.03903",
    archivePrefix = "arXiv",
    primaryClass = "astro-ph.CO",
    doi = "10.1016/j.physrep.2023.11.005",
    journal = "Phys. Rept.",
    volume = "1054",
    pages = "1--68",
    year = "2024"
}

@article{Garcia-Bellido:2024yaz,
    author = "Garcia-Bellido, Juan and Hawkins, Michael",
    title = "{Reanalysis of the MACHO Constraints on PBH in the Light of Gaia DR3 Data}",
    eprint = "2402.00212",
    archivePrefix = "arXiv",
    primaryClass = "astro-ph.GA",
    reportNumber = "IFT-UAM/CSIC-24-14",
    doi = "10.3390/universe10120449",
    journal = "Universe",
    volume = "10",
    number = "12",
    pages = "449",
    year = "2024"
}

@article{Jedamzik:2009uy,
    author = "Jedamzik, Karsten and Pospelov, Maxim",
    title = "{Big Bang Nucleosynthesis and Particle Dark Matter}",
    eprint = "0906.2087",
    archivePrefix = "arXiv",
    primaryClass = "hep-ph",
    doi = "10.1088/1367-2630/11/10/105028",
    journal = "New J. Phys.",
    volume = "11",
    pages = "105028",
    year = "2009"
}

@article{Carr:2009jm,
    author = "Carr, B. J. and Kohri, Kazunori and Sendouda, Yuuiti and Yokoyama, Jun'ichi",
    title = "{New cosmological constraints on primordial black holes}",
    eprint = "0912.5297",
    archivePrefix = "arXiv",
    primaryClass = "astro-ph.CO",
    reportNumber = "RESCEU-31-09, TU-852, YITP-09-112",
    doi = "10.1103/PhysRevD.81.104019",
    journal = "Phys. Rev. D",
    volume = "81",
    pages = "104019",
    year = "2010"
}

@article{Pospelov:2010hj,
    author = "Pospelov, Maxim and Pradler, Josef",
    title = "{Big Bang Nucleosynthesis as a Probe of New Physics}",
    eprint = "1011.1054",
    archivePrefix = "arXiv",
    primaryClass = "hep-ph",
    doi = "10.1146/annurev.nucl.012809.104521",
    journal = "Ann. Rev. Nucl. Part. Sci.",
    volume = "60",
    pages = "539--568",
    year = "2010"
}

@article{Padmanabhan:2005es,
    author = "Padmanabhan, Nikhil and Finkbeiner, Douglas P.",
    title = "{Detecting dark matter annihilation with CMB polarization: Signatures and experimental prospects}",
    eprint = "astro-ph/0503486",
    archivePrefix = "arXiv",
    doi = "10.1103/PhysRevD.72.023508",
    journal = "Phys. Rev. D",
    volume = "72",
    pages = "023508",
    year = "2005"
}

@article{Slatyer:2009yq,
    author = "Slatyer, Tracy R. and Padmanabhan, Nikhil and Finkbeiner, Douglas P.",
    title = "{CMB Constraints on WIMP Annihilation: Energy Absorption During the Recombination Epoch}",
    eprint = "0906.1197",
    archivePrefix = "arXiv",
    primaryClass = "astro-ph.CO",
    doi = "10.1103/PhysRevD.80.043526",
    journal = "Phys. Rev. D",
    volume = "80",
    pages = "043526",
    year = "2009"
}

@article{Chluba:2011hw,
    author = "Chluba, J. and Sunyaev, R. A.",
    title = "{The evolution of CMB spectral distortions in the early Universe}",
    eprint = "1109.6552",
    archivePrefix = "arXiv",
    primaryClass = "astro-ph.CO",
    doi = "10.1111/j.1365-2966.2011.19786.x",
    journal = "Mon. Not. Roy. Astron. Soc.",
    volume = "419",
    pages = "1294--1314",
    year = "2012"
}

@article{Kawasaki:1994af,
    author = "Kawasaki, M. and Moroi, T.",
    title = "{Gravitino production in the inflationary universe and the effects on big bang nucleosynthesis}",
    eprint = "hep-ph/9403364",
    archivePrefix = "arXiv",
    reportNumber = "ICRR-315-94-10, TU-457",
    doi = "10.1143/PTP.93.879",
    journal = "Prog. Theor. Phys.",
    volume = "93",
    pages = "879--900",
    year = "1995"
}

@article{Kawasaki:2017bqm,
    author = "Kawasaki, Masahiro and Kohri, Kazunori and Moroi, Takeo and Takaesu, Yoshitaro",
    title = "{Revisiting Big-Bang Nucleosynthesis Constraints on Long-Lived Decaying Particles}",
    eprint = "1709.01211",
    archivePrefix = "arXiv",
    primaryClass = "hep-ph",
    reportNumber = "KEK-COSMO-211, IPMU17-0117, UT-17-29, KEK-Cosmo-211, KEK-TH-1998",
    doi = "10.1103/PhysRevD.97.023502",
    journal = "Phys. Rev. D",
    volume = "97",
    number = "2",
    pages = "023502",
    year = "2018"
}

@article{Acharya:2019uba,
    author = "Acharya, Sandeep Kumar and Khatri, Rishi",
    title = "{CMB anisotropy and BBN constraints on pre-recombination decay of dark matter to visible particles}",
    eprint = "1910.06272",
    archivePrefix = "arXiv",
    primaryClass = "astro-ph.CO",
    doi = "10.1088/1475-7516/2019/12/046",
    journal = "JCAP",
    volume = "12",
    pages = "046",
    year = "2019"
}

@article{Slatyer:2016qyl,
    author = "Slatyer, Tracy R. and Wu, Chih-Liang",
    title = "{General Constraints on Dark Matter Decay from the Cosmic Microwave Background}",
    eprint = "1610.06933",
    archivePrefix = "arXiv",
    primaryClass = "astro-ph.CO",
    reportNumber = "MIT-CTP-4842",
    doi = "10.1103/PhysRevD.95.023010",
    journal = "Phys. Rev. D",
    volume = "95",
    number = "2",
    pages = "023010",
    year = "2017"
}

@article{Poulin:2016anj,
    author = "Poulin, Vivian and Lesgourgues, Julien and Serpico, Pasquale D.",
    title = "{Cosmological constraints on exotic injection of electromagnetic energy}",
    eprint = "1610.10051",
    archivePrefix = "arXiv",
    primaryClass = "astro-ph.CO",
    doi = "10.1088/1475-7516/2017/03/043",
    journal = "JCAP",
    volume = "03",
    pages = "043",
    year = "2017"
}

@article{Thoss:2024hsr,
    author = "Thoss, Valentin and Burkert, Andreas and Kohri, Kazunori",
    title = "{Breakdown of hawking evaporation opens new mass window for primordial black holes as dark matter candidate}",
    eprint = "2402.17823",
    archivePrefix = "arXiv",
    primaryClass = "astro-ph.CO",
    reportNumber = "KEK-TH-2605;KEK-Cosmo-0339;KEK-QUP-2024-0003",
    doi = "10.1093/mnras/stae1098",
    journal = "Mon. Not. Roy. Astron. Soc.",
    volume = "532",
    number = "1",
    pages = "451--459",
    year = "2024"
}

@article{Alexandre:2024nuo,
    author = "Alexandre, Ana and Dvali, Gia and Koutsangelas, Emmanouil",
    title = "{New mass window for primordial black holes as dark matter from the memory burden effect}",
    eprint = "2402.14069",
    archivePrefix = "arXiv",
    primaryClass = "hep-ph",
    doi = "10.1103/PhysRevD.110.036004",
    journal = "Phys. Rev. D",
    volume = "110",
    number = "3",
    pages = "036004",
    year = "2024"
}

@article{Carr:2020gox,
    author = "Carr, Bernard and Kohri, Kazunori and Sendouda, Yuuiti and Yokoyama, Jun'ichi",
    title = "{Constraints on primordial black holes}",
    eprint = "2002.12778",
    archivePrefix = "arXiv",
    primaryClass = "astro-ph.CO",
    reportNumber = "RESCEU-03/20; KEK-Cosmo-249; KEK-TH-2199; IPMU20-0024",
    doi = "10.1088/1361-6633/ac1e31",
    journal = "Rept. Prog. Phys.",
    volume = "84",
    number = "11",
    pages = "116902",
    year = "2021"
}

@article{Dvali:2024hsb,
    author = "Dvali, Gia and Valbuena-Berm\'udez, Juan Sebasti\'an and Zantedeschi, Michael",
    title = "{Memory burden effect in black holes and solitons: Implications for PBH}",
    eprint = "2405.13117",
    archivePrefix = "arXiv",
    primaryClass = "hep-th",
    doi = "10.1103/PhysRevD.110.056029",
    journal = "Phys. Rev. D",
    volume = "110",
    number = "5",
    pages = "056029",
    year = "2024"
}

@article{Barman:2024iht,
    author = "Barman, Basabendu and Haque, Md Riajul and Zapata, \'Oscar",
    title = "{Gravitational wave signatures of cogenesis from a burdened PBH}",
    eprint = "2405.15858",
    archivePrefix = "arXiv",
    primaryClass = "astro-ph.CO",
    doi = "10.1088/1475-7516/2024/09/020",
    journal = "JCAP",
    volume = "09",
    pages = "020",
    year = "2024"
}

@article{Kohri:2024qpd,
    author = "Kohri, Kazunori and Terada, Takahiro and Yanagida, Tsutomu T.",
    title = "{Induced gravitational waves probing primordial black hole dark matter with the memory burden effect}",
    eprint = "2409.06365",
    archivePrefix = "arXiv",
    primaryClass = "astro-ph.CO",
    reportNumber = "KEK-TH-2654, KEK-Cosmo-0358",
    doi = "10.1103/PhysRevD.111.063543",
    journal = "Phys. Rev. D",
    volume = "111",
    number = "6",
    pages = "063543",
    year = "2025"
}

@article{Bhaumik:2024qzd,
    author = "Bhaumik, Nilanjandev and Haque, Md Riajul and Jain, Rajeev Kumar and Lewicki, Marek",
    title = "{Memory burden effect mimics reheating signatures on SGWB from ultra-low mass PBH domination}",
    eprint = "2409.04436",
    archivePrefix = "arXiv",
    primaryClass = "astro-ph.CO",
    doi = "10.1007/JHEP10(2024)142",
    journal = "JHEP",
    volume = "10",
    pages = "142",
    year = "2024"
}

@article{Jiang:2024aju,
    author = "Jiang, Yang and Yuan, Chen and Li, Chong-Zhi and Huang, Qing-Guo",
    title = "{Constraints on the primordial black hole abundance through scalar-induced gravitational waves from Advanced LIGO and Virgo's first three observing runs}",
    eprint = "2409.07976",
    archivePrefix = "arXiv",
    primaryClass = "astro-ph.CO",
    doi = "10.1088/1475-7516/2024/12/016",
    journal = "JCAP",
    volume = "12",
    pages = "016",
    year = "2024"
}

@misc{Zantedeschi:2024ram,
    author = "Zantedeschi, Michael and Visinelli, Luca",
    title = "{Ultralight Black Holes as Sources of High-Energy Particles}",
    eprint = "2410.07037",
    archivePrefix = "arXiv",
    primaryClass = "astro-ph.HE",
    month = "10",
    year = "2024"
}

@article{Dvali:2025ktz,
    author = "Dvali, Gia and Zantedeschi, Michael and Zell, Sebastian",
    title = "{Transitioning to Memory Burden: Detectable Small Primordial Black Holes as Dark Matter}",
    eprint = "2503.21740",
    archivePrefix = "arXiv",
    primaryClass = "hep-ph",
    month = "3",
    year = "2025"
}

@article{Keith:2020jww,
    author = "Keith, Celeste and Hooper, Dan and Blinov, Nikita and McDermott, Samuel D.",
    title = "{Constraints on Primordial Black Holes From Big Bang Nucleosynthesis Revisited}",
    eprint = "2006.03608",
    archivePrefix = "arXiv",
    primaryClass = "astro-ph.CO",
    reportNumber = "FERMILAB-PUB-20-224-A",
    doi = "10.1103/PhysRevD.102.103512",
    journal = "Phys. Rev. D",
    volume = "102",
    number = "10",
    pages = "103512",
    year = "2020"
}

@article{MacGibbon:1990zk,
    author = "MacGibbon, J. H. and Webber, B. R.",
    title = "{Quark and gluon jet emission from primordial black holes: The instantaneous spectra}",
    doi = "10.1103/PhysRevD.41.3052",
    journal = "Phys. Rev. D",
    volume = "41",
    pages = "3052--3079",
    year = "1990"
}

@article{MacGibbon:1991tj,
    author = "MacGibbon, Jane H.",
    title = "{Quark and gluon jet emission from primordial black holes. 2. The Lifetime emission}",
    reportNumber = "LHEA-91-001",
    doi = "10.1103/PhysRevD.44.376",
    journal = "Phys. Rev. D",
    volume = "44",
    pages = "376--392",
    year = "1991"
}

@article{Coogan:2020tuf,
    author = "Coogan, Adam and Morrison, Logan and Profumo, Stefano",
    title = "{Direct Detection of Hawking Radiation from Asteroid-Mass Primordial Black Holes}",
    eprint = "2010.04797",
    archivePrefix = "arXiv",
    primaryClass = "astro-ph.CO",
    doi = "10.1103/PhysRevLett.126.171101",
    journal = "Phys. Rev. Lett.",
    volume = "126",
    number = "17",
    pages = "171101",
    year = "2021"
}

@article{Boudaud:2018hqb,
    author = "Boudaud, Mathieu and Cirelli, Marco",
    title = "{Voyager 1 $e^\pm$ Further Constrain Primordial Black Holes as Dark Matter}",
    eprint = "1807.03075",
    archivePrefix = "arXiv",
    primaryClass = "astro-ph.HE",
    doi = "10.1103/PhysRevLett.122.041104",
    journal = "Phys. Rev. Lett.",
    volume = "122",
    number = "4",
    pages = "041104",
    year = "2019"
}

@article{PhysRevD.101.123514,
  title = {$INTEGRAL$ constraints on primordial black holes and particle dark matter},
  author = {Laha, Ranjan and Mu\~noz, Julian B. and Slatyer, Tracy R.},
  journal = {Phys. Rev. D},
  volume = {101},
  issue = {12},
  pages = {123514},
  numpages = {9},
  year = {2020},
  month = {Jun},
  publisher = {American Physical Society},
  doi = {10.1103/PhysRevD.101.123514},
  url = {https://link.aps.org/doi/10.1103/PhysRevD.101.123514}
}

@article{Keith:2021guq,
    author = "Keith, Celeste and Hooper, Dan",
    title = "{511~keV excess and primordial black holes}",
    eprint = "2103.08611",
    archivePrefix = "arXiv",
    primaryClass = "astro-ph.CO",
    reportNumber = "FERMILAB-PUB-21-088-T, FERMILAB-PUB-18-309-A",
    doi = "10.1103/PhysRevD.104.063033",
    journal = "Phys. Rev. D",
    volume = "104",
    number = "6",
    pages = "063033",
    year = "2021"
}

@article{Dasgupta:2019cae,
    author = "Dasgupta, Basudeb and Laha, Ranjan and Ray, Anupam",
    title = "{Neutrino and positron constraints on spinning primordial black hole dark matter}",
    eprint = "1912.01014",
    archivePrefix = "arXiv",
    primaryClass = "hep-ph",
    reportNumber = "CERN-TH-2019-212, TIFR/TH/19-40",
    doi = "10.1103/PhysRevLett.125.101101",
    journal = "Phys. Rev. Lett.",
    volume = "125",
    number = "10",
    pages = "101101",
    year = "2020"
}

@article{Macho:2000nvd,
    author = "Allsman, R. A. and others",
    collaboration = "Macho",
    title = "{MACHO project limits on black hole dark matter in the 1-30 solar mass range}",
    eprint = "astro-ph/0011506",
    archivePrefix = "arXiv",
    doi = "10.1086/319636",
    journal = "Astrophys. J. Lett.",
    volume = "550",
    pages = "L169",
    year = "2001"
}

@article{EROS-2:2006ryy,
    author = "Tisserand, P. and others",
    collaboration = "EROS-2",
    title = "{Limits on the Macho Content of the Galactic Halo from the EROS-2 Survey of the Magellanic Clouds}",
    eprint = "astro-ph/0607207",
    archivePrefix = "arXiv",
    doi = "10.1051/0004-6361:20066017",
    journal = "Astron. Astrophys.",
    volume = "469",
    pages = "387--404",
    year = "2007"
}

@article{Griest:2013aaa,
    author = "Griest, Kim and Cieplak, Agnieszka M. and Lehner, Matthew J.",
    title = "{Experimental Limits on Primordial Black Hole Dark Matter from the First 2 yr of Kepler Data}",
    eprint = "1307.5798",
    archivePrefix = "arXiv",
    primaryClass = "astro-ph.CO",
    doi = "10.1088/0004-637X/786/2/158",
    journal = "Astrophys. J.",
    volume = "786",
    number = "2",
    pages = "158",
    year = "2014"
}

@article{Oguri:2017ock,
    author = "Oguri, Masamune and Diego, Jose M. and Kaiser, Nick and Kelly, Patrick L. and Broadhurst, Tom",
    title = "{Understanding caustic crossings in giant arcs: characteristic scales, event rates, and constraints on compact dark matter}",
    eprint = "1710.00148",
    archivePrefix = "arXiv",
    primaryClass = "astro-ph.CO",
    doi = "10.1103/PhysRevD.97.023518",
    journal = "Phys. Rev. D",
    volume = "97",
    number = "2",
    pages = "023518",
    year = "2018"
}

@article{Niikura:2019kqi,
    author = "Niikura, Hiroko and Takada, Masahiro and Yokoyama, Shuichiro and Sumi, Takahiro and Masaki, Shogo",
    title = "{Constraints on Earth-mass primordial black holes from OGLE 5-year microlensing events}",
    eprint = "1901.07120",
    archivePrefix = "arXiv",
    primaryClass = "astro-ph.CO",
    doi = "10.1103/PhysRevD.99.083503",
    journal = "Phys. Rev. D",
    volume = "99",
    number = "8",
    pages = "083503",
    year = "2019"
}

@article{Croon:2020ouk,
    author = "Croon, Djuna and McKeen, David and Raj, Nirmal and Wang, Zihui",
    title = "{Subaru-HSC through a different lens: Microlensing by extended dark matter structures}",
    eprint = "2007.12697",
    archivePrefix = "arXiv",
    primaryClass = "astro-ph.CO",
    doi = "10.1103/PhysRevD.102.083021",
    journal = "Phys. Rev. D",
    volume = "102",
    number = "8",
    pages = "083021",
    year = "2020"
}

@article{Zeldovich:1967lct,
    author = "Zel'dovich, Ya. B. and Novikov, I. D.",
    title = "{The Hypothesis of Cores Retarded during Expansion and the Hot Cosmological Model}",
    journal = "Sov. Astron.",
    volume = "10",
    pages = "602",
    year = "1967"
}

@article{Hawking:1971ei,
    author = "Hawking, Stephen",
    title = "{Gravitationally collapsed objects of very low mass}",
    doi = "10.1093/mnras/152.1.75",
    journal = "Mon. Not. Roy. Astron. Soc.",
    volume = "152",
    pages = "75",
    year = "1971"
}

@article{Chapline:1975ojl,
    author = "Chapline, George F.",
    title = "{Cosmological effects of primordial black holes}",
    doi = "10.1038/253251a0",
    journal = "Nature",
    volume = "253",
    number = "5489",
    pages = "251--252",
    year = "1975"
}

@article{Meszaros:1975ef,
    author = "Meszaros, P.",
    title = "{Primeval black holes and galaxy formation}",
    journal = "Astron. Astrophys.",
    volume = "38",
    pages = "5--13",
    year = "1975"
}

@article{Bekenstein:1973ur,
    author = "Bekenstein, Jacob D.",
    title = "{Black holes and entropy}",
    doi = "10.1103/PhysRevD.7.2333",
    journal = "Phys. Rev. D",
    volume = "7",
    pages = "2333--2346",
    year = "1973"
}

@article{Hawking:1975vcx,
    author = "Hawking, S. W.",
    editor = "Gibbons, G. W. and Hawking, S. W.",
    title = "{Particle Creation by Black Holes}",
    doi = "10.1007/BF02345020",
    journal = "Commun. Math. Phys.",
    volume = "43",
    pages = "199--220",
    year = "1975",
    note = "[Erratum: Commun.Math.Phys. 46, 206 (1976)]"
}

@article{Dvali:2018xpy,
    author = "Dvali, Gia",
    title = "{A Microscopic Model of Holography: Survival by the Burden of Memory}",
    eprint = "1810.02336",
    archivePrefix = "arXiv",
    primaryClass = "hep-th",
    month = "10",
    year = "2018"
}

@article{Dvali:2018ytn,
    author = "Dvali, Gia and Eisemann, Lukas and Michel, Marco and Zell, Sebastian",
    title = "{Universe's Primordial Quantum Memories}",
    eprint = "1812.08749",
    archivePrefix = "arXiv",
    primaryClass = "hep-th",
    reportNumber = "LMU-ASC 82/18; MPP-2018-302",
    doi = "10.1088/1475-7516/2019/03/010",
    journal = "JCAP",
    volume = "03",
    pages = "010",
    year = "2019"
}

@article{Dvali:2020wft,
    author = "Dvali, Gia and Eisemann, Lukas and Michel, Marco and Zell, Sebastian",
    title = "{Black hole metamorphosis and stabilization by memory burden}",
    eprint = "2006.00011",
    archivePrefix = "arXiv",
    primaryClass = "hep-th",
    doi = "10.1103/PhysRevD.102.103523",
    journal = "Phys. Rev. D",
    volume = "102",
    number = "10",
    pages = "103523",
    year = "2020"
}

@article{Dvali:2021tez,
    author = "Dvali, Gia and Kaikov, Oleg and Berm\'udez, Juan Sebasti\'an Valbuena",
    title = "{How special are black holes? Correspondence with objects saturating unitarity bounds in generic theories}",
    eprint = "2112.00551",
    archivePrefix = "arXiv",
    primaryClass = "hep-th",
    doi = "10.1103/PhysRevD.105.056013",
    journal = "Phys. Rev. D",
    volume = "105",
    number = "5",
    pages = "056013",
    year = "2022"
}

@article{Dvali:2023xfz,
    author = "Dvali, Gia",
    title = "{Saturon Dark Matter}",
    eprint = "2302.08353",
    archivePrefix = "arXiv",
    primaryClass = "hep-ph",
    month = "2",
    year = "2023"
}

@article{Michel:2023ydf,
    author = "Michel, Marco and Zell, Sebastian",
    title = "{The Timescales of Quantum Breaking}",
    eprint = "2306.09410",
    archivePrefix = "arXiv",
    primaryClass = "quant-ph",
    doi = "10.1002/prop.202300163",
    journal = "Fortsch. Phys.",
    volume = "71",
    pages = "2300163",
    year = "2023"
}

@article{Dvali:2017ktv,
    author = "Dvali, Gia",
    title = "{Critically excited states with enhanced memory and pattern recognition capacities in quantum brain networks: Lesson from black holes}",
    eprint = "1711.09079",
    archivePrefix = "arXiv",
    primaryClass = "quant-ph",
    month = "11",
    year = "2017"
}

@article{Dvali:2018tqi,
    author = "Dvali, Gia and Michel, Marco and Zell, Sebastian",
    title = "{Finding Critical States of Enhanced Memory Capacity in Attractive Cold Bosons}",
    eprint = "1805.10292",
    archivePrefix = "arXiv",
    primaryClass = "quant-ph",
    reportNumber = "LMU-ASC 31/18; MPP-2018-112, LMU-ASC-31-18, MPP-2018-112",
    doi = "10.1140/epjqt/s40507-019-0071-1",
    journal = "EPJ Quant. Technol.",
    volume = "6",
    pages = "1",
    year = "2019"
}

\end{document}